\documentclass[prd,aps,twocolumn]{revtex4}

\usepackage{epsfig,graphicx,tabularx}
\usepackage{amsmath,amssymb,mathrsfs}
\usepackage{psfrag}
\usepackage{graphics}
\usepackage{xcolor}
\usepackage{bm}
\usepackage{verbatim,ulem}
\usepackage{hyperref}

\newcommand{\beq}{\begin{equation}}
\newcommand{\eeq}{\end{equation}}
\newcommand{\bal}{\begin{aligned}[b]}
\newcommand{\eal}{\end{aligned}}
\newcommand{\beqa}{\begin{eqnarray}}
\newcommand{\eeqa}{\end{eqnarray}}

\newcommand{\rev}[1]{\textcolor{black}{#1}}

\begin{document}

  \title{Gaussian fluctuations in the tunneling probability \\ of a
    closed universe} 

\author{Luca Salasnich} 

\affiliation{Dipartimento di Fisica e Astronomia
    'Galileo Galilei' and Padua QTech Center, Universit\`a di Padova,
            via Marzolo 8, Padova, 35131, Italy}

\affiliation{Istituto Nazionale di Fisica Nucleare,
    Sezione di Padova, via Marzolo 8, Padova, 35131, Italy}

\begin{abstract}
  We consider the quantum creation of a closed universe within the Euclidean
  path-integral formalism.
  An analytical expression for the tunneling probability is derived,
  including both the exponential suppression and the exact Gaussian prefactor
  due to quadratic fluctuations around the instanton.
  \rev{The calculation is performed in a fixed-interval minisuperspace
  formulation, where the Hamiltonian constraint is imposed at the level of
  the classical instanton, while the full lapse integration is not included
  beyond the leading semiclassical approximation.}
  The result provides
  a transparent and self-consistent semiclassical estimate of the nucleation
  rate, refining previous analyses with the inclusion of
  Gaussian fluctuations.
\end{abstract}

\maketitle

\section{Introduction}

Understanding the quantum origin of the universe remains one of the
fundamental challenges in theoretical cosmology \cite{cosmo-book}.  
In the framework of closed Friedmann-Lemaitre-Robertson-Walker (FLRW)
models, the nucleation of a finite-size universe can be described as a
quantum tunneling process from ``nothing'', that is from the classically
forbidden region where the scale factor vanishes \cite{pagels,vilenkin}.
This scenario provides a concrete and physically motivated mechanism
for the quantum creation of the universe.
The Euclidean path integral offers a natural framework to describe such
a process, expressing the transition amplitude between two
three-geometries as a sum over all interpolating Euclidean metrics.
In this approach, the semiclassical tunneling probability is dominated
by the contribution of the instanton solution that extremizes the
Euclidean action, leading to an exponential term
governed by the cosmological constant. This picture was first formulated in the
seminal works of Atkatz and Pagels \cite{pagels} and
Vilenkin \cite{vilenkin}. Quite remarkably, although the leading exponential
dependence of the tunneling probability is well understood 
\cite{pagels,vilenkin,linde,zeldovich,rubakov},  
and also generalized including massive scalar
fields \cite{vilenkin2,vilenkin2018,jia2023,lehners}, 
the accurate determination of the prefactor due to Gaussian fluctuations
was never performed exactly, because it requires handling a
complicated differential operator with non-constant coefficients
and carefully treating boundary singularities in the instanton solution.
\rev{To the best of our knowledge, an explicit closed analytical expression
for the Gaussian fluctuation prefactor in this Euclidean closed-FLRW
 tunneling problem has not been previously derived. The aim of the present
work is therefore not to formulate a complete constrained gravitational
path integral, but to isolate and evaluate exactly the quadratic fluctuation
determinant associated with the standard semiclassical Euclidean instanton.}

\begin{figure}[t]
\centerline{\epsfig{file=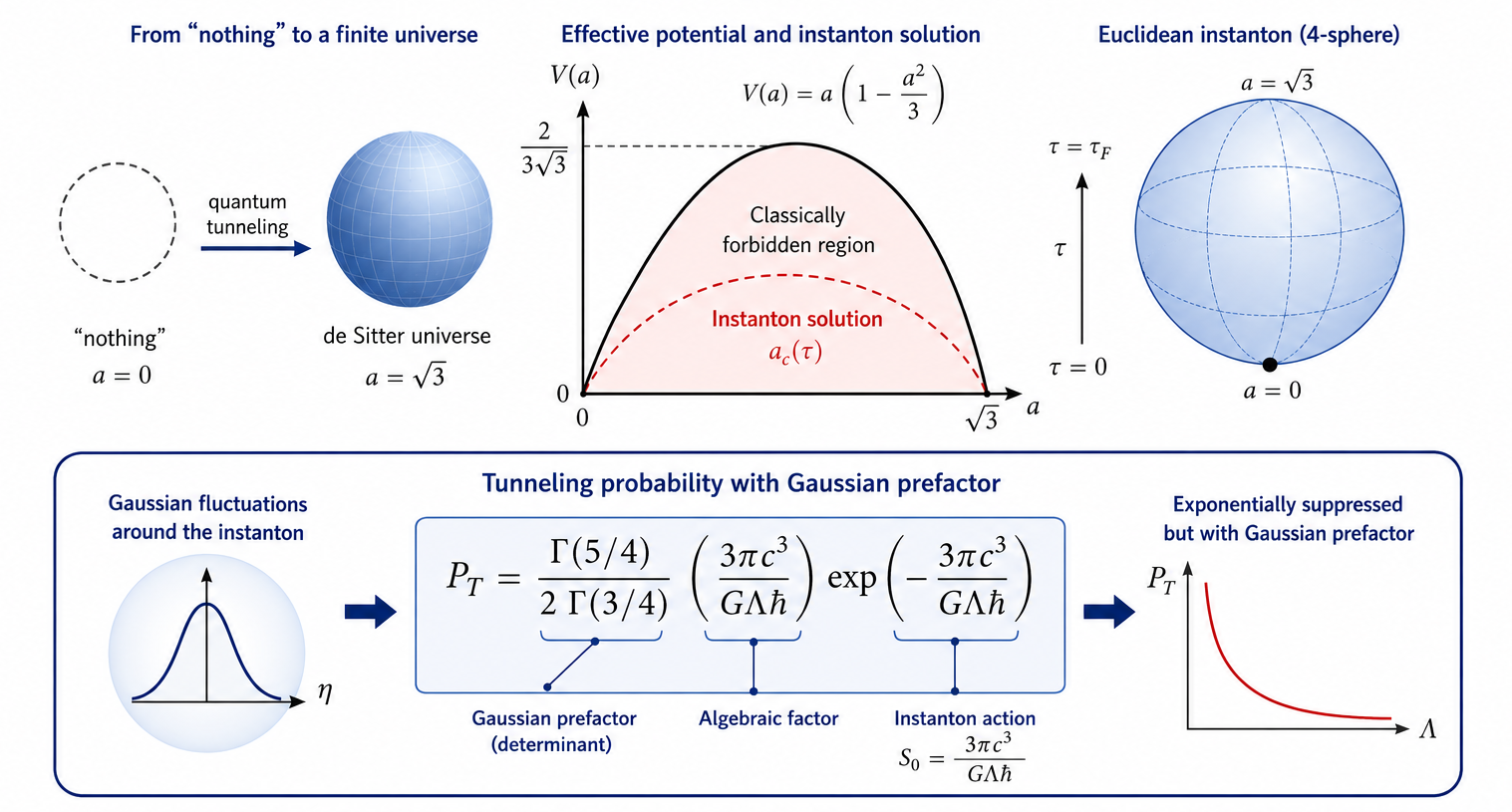,width=9cm,clip=}}
\small 
\caption{Graphical summary of the main results discussed in this paper:
  the analytical calculation of the quantum tunneling probability
  of a closed universe including Gaussian fluctuations.} 
\label{fig1}
\end{figure} 

In this work we adopt the path-integral formulation of quantum
cosmology for a closed FLRW universe and derive a fully analytical 
expression for the tunneling probability, including both the exponential
suppression and the exact Gaussian prefactor arising from quadratic
fluctuations around the instanton solution. See also Fig. \ref{fig1}.
Our derivation extends the 
analyses of Refs. \cite{pagels,vilenkin}, yielding a transparent 
and self-consistent semiclassical estimate of the tunneling rate, and
clarifying the role of the fluctuation determinant in the Euclidean path
integral for quantum cosmology.
It is important to emphasize that the issue of fluctuations
  in quantum cosmology has been discussed in detail in several works
  in the literature. 
  For example, Ref.~\cite{dealwis2019} analyzes the nature of quadratic
  fluctuations around homogeneous backgrounds within a
  Hamiltonian/Wheeler-DeWitt formalism, including implications for
  cosmological perturbations. 
  Ref.~\cite{halliwell2019} investigates conceptual aspects of the
  no-boundary wave function and the structure of fluctuations in that context,
  focusing on the definition of the quantum state rather than on an
  explicit evaluation of a fluctuation determinant.
  Moreover, Ref.~\cite{feldbrugge2017} explores the Lorentzian path integral
  approach with complex contour deformations as a foundation for quantum
  cosmology. \rev{More recent developments have further investigated
  Lorentzian and Picard-Lefschetz formulations of quantum cosmology,
  including truly Lorentzian minisuperspace path integrals, resurgence
  methods around cosmological saddle points, and extensions involving
  torsion \cite{jia2023,honda2024,mondal2024}.}
  In contrast, the present work remains within the standard
  Euclidean tunneling proposal for a closed FLRW universe and focuses on the
  explicit analytical evaluation of the Gaussian prefactor associated with
  quadratic fluctuations around the Euclidean instanton solution.
  This provides a closed-form expression for the semiclassical tunneling
  probability and quantifies directly the contribution of quadratic fluctuations
  to the prefactor, without relying on contour deformations or numerical
  approximations.

\section{Closed Friedmann-Lemaitre-Robertson-Walker universe}

Let us consider the action functional of a closed
Friedmann-Lemaitre-Robertson-Walker universe \cite{cosmo-book} 
\beq
S[a(t)] = - A \int dt \, [ a \, {\dot a}^2 - V(a) ] \; , 
\label{actiona}
\eeq
where
\beq
A = {3\pi c^3\over 4G\Lambda}  
\eeq
with $c$ the speed of light in vacuum, $G$ the gravitational constant, and 
$\Lambda$ the cosmological constant. Here $a(t)$ is the scale
factor of this minisuperspace approximation of the universe with 
\beq
V(a) = a \left( 1 - {1\over 3}a^2 \right) 
\eeq
a sort of effective potential energy, and ${\dot a}=da/dt$. Please note that
$a$ is dimensionless and it must be multiplied by $a_0=1/\sqrt{\Lambda}$
to get the dimensional one. Similarly, $t$ is the dimensionless time and 
it must be multiplied by $t_0=a_0/c$ to get the dimensional one. 
The constant of motion during the time evolution is 
\beq
E =  -A \, [  a \, {\dot a}^2 + V(a) ] \; . 
\eeq
Notice that both $A$ and $E$ have the dimensional units of an action
(Joule $\times$ seconds). In quantum cosmology the lapse
function is the Lagrange multiplier enforcing the Hamiltonian constraint
and reflects the invariance under time reparametrizations. 
\rev{In the present reduced formulation we work in a fixed proper-time
gauge, so that the lapse is not treated as an independent dynamical
variable in the fluctuation determinant. Equivalently, the zero-energy
Hamiltonian constraint is imposed on the classical instanton solution,
while the Gaussian determinant is evaluated for fluctuations of the scale
factor with fixed endpoints over the corresponding Euclidean time interval.}
This allows us to compute Gaussian fluctuations
directly for this fixed-interval system, providing a well-defined
semiclassical prefactor \rev{within the chosen gauge-fixed minisuperspace
setting, although it does not represent the full Faddeev-Popov or BRST
quantization of the gravitational path integral.}

\section{Quantum tunneling from ``nothing''}

In the description of a closed FLRW universe, invariance under
time reparametrization leads to the constraint $E=0$ \cite{vilenkin,dewitt}.
Under this constraint the region with $0\leq a\leq \sqrt{3}$ is
classically forbidden.
One can calculate the probability rate $P_T$ of quantum tunneling from 
$a=0$ to $a=\sqrt{3}$ by using the quantum propagator 
$\langle a_F=\sqrt{3},t_F|a_I=0,t_I=0\rangle$ as follows
\beq
P_T = |\langle a_F=\sqrt{3},t_F|a_I=0,t_I=0\rangle|^2 \; . 
\eeq
The ratio is needed to have a dimensionless quantity with
initial probability $P_T$ equal to one. 
The quantum propagator can be written as a Feynman
path integral \cite{hibbs-book,altland-book}
\beq
\langle a_F=\sqrt{3},t_F|a_I=0,t_I=0\rangle =
\int {\cal D}[a(t)] \, e^{{i\over\hbar}S[a(t)]} \; . 
\eeq
Performing a Wick rotation
\beq
t = i \tau 
\eeq
we have
\beq
\langle a_F=\sqrt{3},\tau_F|a_I=0,\tau_I=0 \rangle =
\int {\cal D}[a(\tau)] \, e^{-S_E[a(\tau)]/\hbar}
\eeq
with
\beq
S_E[a(\tau)] = A \int_0^{\tau_F}
d\tau \, [ a \, {a'}^2 + V(a) ]
\label{euclidean}
\eeq
the Euclidean action and $a'=da/d\tau$. Quite remarkably
\beq
a_c(\tau) = \sqrt{3} \, \sin\left({\tau\over \sqrt{3}}\right)
\label{ac}
\eeq
is the instanton solution that
extremizes $S_E[a(\tau)]$ with boundary conditions
$a_c(0)=0$ and $a_c(\tau_F)=\sqrt{3}$ provided that
\beq
\tau_F = {\pi \over 2} \sqrt{3} \; .
\label{tauf}
\eeq
The Euclidean action for this solution can be evaluated in closed
form: 
\beq
S_E[a_c(\tau)] = 2 A = {3\pi c^3\over 2G\Lambda} \; .
\label{ausful1}
\eeq
It follows immediately that within the saddle point approximation,
where
\beq
P_T = |e^{-S_E[a_c(\tau)]/\hbar}|^2 \; , 
\eeq
the tunneling probability reads 
\beq
P_T = \exp{\left( -{3\pi c^3\over G\Lambda \hbar} \right)} \; .
\label{simplePT}
\eeq
This result was obtained for the first time by Atkatz and Pagels 
\cite{pagels} and Vilenkin \cite{vilenkin} in 1982, see also
Refs. \cite{linde,zeldovich,rubakov,vilenkin2,vilenkin2018}.
The exponentially decreasing behavior obtained here arises 
from the Wick rotation $t = i\tau$, that is the Vilenkin tunneling proposal.
In the Hartle-Hawking no-boundary proposal \cite{hartle}, 
the opposite sign in the exponent originates from the Wick
rotation $t=-i\tau$ \cite{linde-book}, leading to an opposite sign
    in the exponential and a wave function that favors smaller
    initial universes, in contrast with the Vilenkin tunneling approach
    considered here, which tends to select initial expansion.
The question of which proposal is the correct approach is still 
debated in quantum cosmology, and there is no definitive 
consensus \cite{lehners}. However, the Vilenkin tunneling proposal 
generally leads to initial conditions that make inflation
    more probable, since a sufficiently large initial value of the cosmological
constant $\Lambda$ is required to drive an inflationary phase,
whereas small values of $\Lambda$ would not lead to inflation
and are exponentially suppressed in the tunneling
probability \cite{vilenkin2018,linde-book}.
\rev{Recent analyses of modified tunneling wave functions and Lorentzian
path integrals show that this question remains active, especially when
quantum-gravity corrections or different integration contours are included
\cite{motaharfar2023,honda2024,jia2023}.}

\section{Gaussian fluctuations}

Before proceeding with the calculation of Gaussian fluctuations,
  it is important to note that, in a full quantum gravitational treatment,
  the wave function of the universe $\psi$ must satisfy the Hamiltonian
  constraint ${\hat H} \psi = 0$, that is the Wheeler-DeWitt
  equation \cite{dewitt}, which encodes time-reparametrization invariance.
  Expanding the Euclidean action around the classical instanton and computing
  the prefactor directly, as done in the present work, does not strictly
  enforce this constraint beyond leading semiclassical order. 
  Nevertheless, within the minisuperspace approximation and the standard
  Euclidean tunneling approach, this procedure provides a well-defined and
  fully analytical estimate of the contribution of quadratic fluctuations
  to the semiclassical tunneling probability. From this point of view, 
 our strategy is consistent with previous semiclassical treatments
  in the literature \cite{pagels,vilenkin,linde}, and offers a simplified, 
  but analytically tractable, perspective compared to approaches
  \cite{dealwis2019, halliwell2019} that enforce
  time-reparametrization invariance at the quantum level.

\rev{We also stress that the Euclidean formulation of quantum gravity is
known to suffer from the conformal-factor problem, namely from the fact
that the Euclidean gravitational action is not bounded from below in the
space of all metrics. The present minisuperspace calculation does not aim
to solve this general problem. Rather, it should be understood as a
semiclassical calculation performed around the closed-FLRW tunneling
instanton, in the same spirit as the original Euclidean tunneling
treatments. Alternative Lorentzian contour formulations provide a more
fundamental way of addressing some of these issues, but they lead to a
different path-integral prescription and are not the framework adopted
here \cite{feldbrugge2017,jia2023,honda2024,mondal2024}.}

Let us now expand quadratically the Euclidean action $S_E[a(\tau)]$ around
the instanton classical solution $a_c(\tau)$ using
\beq
a(\tau) = a_c(\tau) + \delta a(\tau)
\eeq
with $\delta a(\tau)$ representing a small fluctuation such that 
$\delta a(0)=\delta a(\tau_F)=0$, i.e. with Dirichlet boundary conditions.
In particular, the quadratic contribution from
the kinetic term reads $a_c (\delta a')^2 + 2 a_c' \delta a \delta a'$,
while the potential contributes $\frac12 V''(a_c) (\delta a)^2$.  
The mixed term $2 a_c' \delta a \delta a'$ can be integrated by parts,
and the term $a_c (\delta a')^2$ can be written in symmetric form using
integration by parts. After these manipulations we find 
\beqa
S_E[a(\tau)] &=& S_E[a_c(\tau)]
\\
&+& {1\over 2} \int_0^{\tau_F} d\tau \int_0^{\tau_F}
d\tau' \delta a(\tau') \, {\delta^2 S_E\over \delta a(\tau) \delta a(\tau')}
\, \delta a(\tau)  \; ,
\nonumber
\eeqa
where
\beqa
{\delta^2 S_E\over \delta a(\tau) \delta a(\tau')}
&=& \delta(\tau -\tau') \, \Big[ 
  - 2A {d\over d\tau}
  \left( a_c(\tau) {d\over d\tau}\right)
  \nonumber 
\\  
&+& A \left( V''(a_c(\tau)) - 2 a_c''(\tau) \right) \Big] \; .
\label{S2tox}
\eeqa
Including these Gaussian fluctuations the propagator 
becomes \cite{hibbs-book,altland-book}
\beq
\langle a_F=\sqrt{3},\tau_F|a_I=0,\tau_I=0\rangle = F_E \, 
e^{-S_E[a_c(\tau)]/\hbar} \;
\eeq
where
\beq
F_E = {\cal C} \,
{\rm det}\left[{1\over \hbar}
  {\delta^2 S_E\over \delta a(\tau) \delta a(\tau')} \right]^{-1/2}
\label{rivermystic}
\eeq
with ${\cal C}$ the appropriate (infinite) constant
related to the measure of the path integral.

In contrast to ordinary vacuum decay in quantum field theory, where
the Euclidean fluctuation operator has one negative eigenvalue leading
to an imaginary part of the energy
\cite{coleman1977,callancoleman}, the present model describes a
fixed-endpoint tunneling amplitude in minisuperspace rather than a decay
rate per unit spacetime volume. 
\rev{The Dirichlet boundary conditions remove the translational zero mode
that is present in bounce calculations with time-translation invariance.
Moreover, possible sign ambiguities of the reduced fluctuation determinant
do not have the same interpretation as in false-vacuum decay, because the
quantity computed here is not an imaginary part of an energy but the
absolute semiclassical tunneling probability between two fixed boundary
geometries.}
For this reason we take the absolute value of the determinant of quadratic
fluctuations, ensuring that the prefactor remains real and positive. 
With this prescription,
\beq 
P_T = |F_E|^2 \, e^{-2S_E/\hbar}
\label{PTuseful}
\eeq
represents the absolute semiclassical probability for tunneling
in this minisuperspace, rather than a real-time decay rate. 

\subsection{Approximate prefactor}

To compute the prefactor $F_E$ of Eq. (\ref{PTuseful})
we approximate the classical bounce $a_c(\tau)$
near the barrier maximum (inverted potential minimum)
$a_m = \sqrt{3}/2$, where the Euclidean action 
can be expanded quadratically around $a_m$. 
Namely, for the calculation of Gaussian fluctuations
  instead of (\ref{euclidean}) we use the simplified action 
\beq
S_E[a(\tau)] = A \int_0^{\tau_F} d\tau \left[ a_m {a'}^2 + V(a_m)
+ {1\over 2} V''(a_m) (a - a_m)^2 \right] 
\eeq
with $V''(a_m)=-\sqrt{3}$. In this approximation we get 
\beq
{\delta^2 S_E\over \delta a(\tau) \delta a(\tau')}
= \delta(\tau -\tau') \,
\Big[ - \sqrt{3} A {d^2\over d\tau^2} - \sqrt{3} A \Big] \; . 
\eeq
The Euclidean propagator for this quadratic problem is exactly that of an
inverted harmonic oscillator of effective mass $m= \sqrt{3} A$ and effective
frequency $\omega=1$. For the harmonic
oscillator of frequency $\omega$, in real time
the prefactor of the propagator is \cite{hibbs-book,altland-book}
\beq
F = \sqrt{m\omega\over 2\pi i \hbar \sin(\omega t_F)} \; . 
\eeq
For the Euclidean (imaginary time) inverted harmonic oscillator 
we have $t_F=-i \tau_F$ and consequently, taking into account
that $\sin(-ix)=-i\sinh(x)$, the Euclidean prefactor is 
\beq
F_E = \sqrt{m\omega\over 2\pi \hbar \sinh(\omega \tau_F)} \; . 
\eeq
In particular, for our problem with $m=\sqrt{3}A$, $\omega=1$, and
  $\tau_F=\pi\sqrt{3}/2$ we obtain 
\beq
F_E = \sqrt{\sqrt{3} A
  \over 2\pi \hbar \sinh({\sqrt{3}\pi\over 2})} \; .
\label{ausful2}
\eeq
Thus, by using Eqs. (\ref{ausful1}) and (\ref{ausful2})
the tunneling probability (\ref{PTuseful}) is given by 
\beqa
P_T &=& {\sqrt{3}\over 4\sinh(\pi \sqrt{3}/2)} \,
\left({3\pi c^3\over G\Lambda \hbar}\right) 
\exp{\left( -{3\pi c^3\over G\Lambda \hbar} \right)}
\nonumber
\\
&\simeq& 0.018 \, \left({3\pi c^3\over G\Lambda \hbar}\right) 
\exp{\left( -{3\pi c^3\over G\Lambda \hbar} \right)} \, .
\label{approxPT}
\eeqa

\subsection{Exact prefactor}

To obtain an exact formula for the prefactor of the Gaussian fluctuations
term we notice that Eq. (\ref{S2tox}) can be re-written as
\beq
{\delta^2 S_E\over \delta a(\tau) \delta a(\tau')}
= \delta(\tau -\tau') \, 2 A \, {\cal L} \; ,
\eeq
where 
\beq 
{\cal L} = \Big[ 
  - {d\over d\tau}
  \left( a_c(\tau) {d\over d\tau}\right) 
  + {1\over 2} \left( V''(a_c(\tau)) - 2 a_c''(\tau) \right) \Big]
\label{mino0}
\eeq
is a quite nasty differential operator. Our strategy is to
derive another operator, ${\tilde {\cal L}}$, such that
${\rm det}[{\cal L}]={\rm det}[{\tilde {\cal L}}]$. 
The new operator ${\tilde {\cal L}}$ will be simpler and for it
we can use the Gel'fand-Yaglom theorem \cite{gy-method1,gy-method2},
which also helps us to fix the constant ${\cal C}$ which appears
in the definition of $F_E$, see Eq. (\ref{rivermystic}).
This procedure, discussed in the Appendix, gives
the exact Euclidean prefactor of Gaussian fluctuations:  
\beq
F_E = \sqrt{\frac{A}{2\pi \hbar}} \,
\frac{\Gamma(5/4)}{\Gamma(3/4)} \; .
\label{bastacosi}
\eeq
Therefore, recalling Eqs. (\ref{ausful1}) and (\ref{PTuseful}) 
the tunneling probability reads 
\beqa
P_T &=&  {\Gamma(5/4)^2\over 2 \Gamma(3/4)^2} \, 
\left( \frac{3\pi c^3}{G\Lambda \hbar}\right) \, 
\exp\Big(-\frac{3\pi c^3}{G\Lambda \hbar}\Big) 
\nonumber 
\\
&\simeq& 0.318 \, 
\left( \frac{3\pi c^3}{G\Lambda \hbar}\right) \, 
\exp\Big(-\frac{3\pi c^3}{G\Lambda \hbar}\Big) \; .
\label{betterPT}
\eeqa
This result provides a meaningful evaluation of the quantum tunneling
probability, improving on the estimate obtained using the inverted harmonic
oscillator approximation, Eq. (\ref{approxPT}). Notice that 
the difference between Eqs. (\ref{approxPT}) and (\ref{betterPT}) is only
in the numerical value of the prefactor. This is not fully surprising: 
the dominant contribution to the fluctuation determinant comes from the
region near the instanton peak, where the potential is accurately described
by the harmonic approximation.

The validity of the semiclassical approximation employed here
    relies on the saddle-point expansion of the Euclidean path integral.
    In the present model this requires the Euclidean action of
    the instanton to be large compared to $\hbar$, which translates into
    the condition $3\pi c^3/(2G\Lambda\hbar) \gg 1$. Under this assumption,
    higher-order (non-Gaussian) fluctuations are parametrically suppressed,
    and the Gaussian prefactor computed here provides the leading
    quantum correction to the tunneling probability. \rev{The prefactor
    changes the leading result only by an algebraic factor, while the
    dominant dependence on the cosmological constant remains exponentially
    controlled by the instanton action. Thus the correction is important
    for a consistent semiclassical normalization, but it does not alter the
    exponential hierarchy of nucleation probabilities.}
    The results
  presented here provide a reliable semiclassical estimate of the tunneling
  probability and the associated prefactor, but do not capture fully
  nonperturbative quantum effects, which could be accessed only
  through numerical integration of the full path integral or direct solution
  of the Wheeler-DeWitt equation. Indeed, one possible beyond-semiclassical
  approach is the direct numerical solution of the
  Wheeler-DeWitt equation in minisuperspace, which fully enforces the
  Hamiltonian constraint and captures nonperturbative quantum
  effects \cite{vilenkin2018,jia2023}. 
  Alternatively, one could try a numerical evaluation of the Euclidean
  path integral using
  discretization, lattice techniques, or Monte Carlo methods can account for
  higher-order fluctuations beyond the Gaussian level. 
Among these numerical alternatives, a direct numerical solution of the
Wheeler-DeWitt equation appears more suitable for the present
one-dimensional minisuperspace model, since it implements the
Hamiltonian constraint exactly and reduces the problem to a
well-defined ordinary differential equation. A numerical evaluation
of the path integral would be technically more involved and would not
provide additional advantages for a system with a single degree of freedom.
However, these numerical methods are typically computationally
intensive and less transparent than the analytical semiclassical approach
presented here.  Thus, the analytical calculation of the Gaussian prefactor
remains valuable, as it gives a clear and tractable estimate of the
first-order quantum corrections to the tunneling probability.

\section{Conclusions}

We have presented an analytical evaluation of the quantum
tunneling probability for a closed FLRW universe, including both the 
exponential suppression and the prefactor due to Gaussian fluctuations 
around the instanton solution. The prefactor has been calculated
in two ways: an approximated procedure based on a harmonic expansion near
the barrier maximum, and a fully analytical calculation employing an
isospectral transformation and the Gel'fand-Yaglom theorem. 
\rev{The main novelty of the present work is the explicit analytical
evaluation of the Gaussian determinant associated with the reduced
Euclidean fluctuation operator. The calculation clarifies how the
quadratic prefactor modifies the leading tunneling probability: it changes
the overall normalization by an algebraic factor proportional to
$3\pi c^3/(G\Lambda\hbar)$, while the dominant dependence remains governed
by the exponential instanton action. The result is therefore meaningful
only in the semiclassical regime $S_E/\hbar \gg 1$ and within the
minisuperspace truncation.}
Beyond the Gaussian approximation, the tunneling probability
could in principle be obtained numerically either from the full Euclidean
path integral, using discretization and Monte Carlo or lattice methods
to capture higher-order quantum fluctuations, or directly by solving the
Wheeler-DeWitt equation in minisuperspace, providing a fully nonperturbative
evaluation of the nucleation probability.
The tunneling probability calculated
  here should be interpreted as a semiclassical measure of the likelihood
  for universe nucleation in minisuperspace. The approach follows the
  Vilenkin tunneling proposal \cite{vilenkin} and differs from the
  Hartle-Hawking no-boundary prescription \cite{hartle}, which would
  invert the sign of the exponential and favor smaller initial universes.
\rev{Finally, although the present study is concerned with quantum
cosmology rather than compact objects, recent investigations of charged
black holes and electromagnetic effects in alternative theories of gravity
provide complementary examples of how strong-gravity systems may be used
to test gravitational dynamics beyond the simplest Einstein-Hilbert setting
\cite{mushtaq2025,kala2022}. These topics are outside the scope of the
present minisuperspace calculation, but they illustrate the broader
relativistic context in which semiclassical and strong-field gravitational
phenomena are currently being explored.}

\section*{Funding}

\rev{This work is partially supported by the National project ``Frontiere
Quantistiche'' (Dipartimenti di Eccellenza MUR 2023--2027) and by
``Iniziativa Specifica Quantum'' of INFN. The APC funding will be specified
at the publication stage, if applicable.}

\section*{Data Availability}

\rev{No datasets were generated or analyzed during the present study.}

\section*{Acknowledgments}

LS thanks Marco Peloso and Alessandro Pennacchio for useful discussions.

\section*{Conflict of Interest}

\rev{The author declares no conflict of interest.}

  \section*{Appendix}

In this Appendix we provide the technical details of the calculation
of the fluctuation determinant entering the prefactor $F_E$.
Starting from the operator ${\cal L}$ of Eq. (\ref{mino0}),
obtained from the second variation
of the Euclidean action, we perform an isospectral transformation to a
Schr\"odinger-like operator ${\tilde {\cal L}}$ and apply the
Gel'fand-Yaglom theorem \cite{gy-method1,gy-method2}
to evaluate the corresponding functional determinant.

Let us consider the operator ${\cal L}$ of Eq. (\ref{mino0}) which
satisfies the differential problem 
\beq
0 = {\cal L} \, \delta a(\tau) \; . 
\eeq
An isospectral transformation
\beq
\delta a(\tau) = \frac{\chi(\tau)}{\sqrt{a_c(\tau)}} 
\eeq
with $a_c(\tau)=\sqrt{3}\sin(\tau/\sqrt{3})$ gives a Schr\"odinger-like
(more precisely, Sturm-Liouville type) differential equation 
\beqa
0 = {\tilde {\cal L}} \, \chi(\tau) = \left[ - \frac{d^2}{d\tau^2} + U(\tau) 
  \right] \chi(\tau) \;
\label{serve1}
\eeqa
with
\beq
U(\tau) = -\frac{3}{4} -
\frac{1}{12} {1\over \sin^2\big(\frac{\tau}{\sqrt{3}}\big)} \; . 
\eeq
This transformation rewrites the original Sturm-Liouville problem
in a Schr\"odinger-like form with standard second-derivative structure,
so that the Gel'fand-Yaglom theorem \cite{gy-method1,gy-method2}
can be applied directly.
Since the transformation is isospectral, the functional determinants
of ${\cal L}$ and ${\tilde {\cal L}}$ coincide.

In our problem the Gel'fand-Yaglom theorem \cite{gy-method1,gy-method2}
implies that 
\beq
F_E = \sqrt{A\over \pi \hbar} \left(
\frac{\det {\tilde {\cal L}}}{\det {\tilde {\cal L}}_0}\right)^{-1/2} =
 \sqrt{A\over \pi \hbar} 
 \left( \frac{\chi(\tau_F)}{\chi_0(\tau_F)} \right)^{-1/2} \, ,
 \label{mino1}
 \eeq
 where the term $\sqrt{A/(\pi\hbar)}$ 
 originates from the path integral measure and from the factor $2A$
 appearing in the second variation of the Euclidean action.
 This ensures that the prefactor has the correct dimension and
 is consistent with the standard semiclassical limit. 
The reference operator ${\tilde {\cal L}}_0$ is introduced solely to define
the determinant ratio appearing in the Gel'fand-Yaglom formula.
Any regular operator with known solution satisfying the same boundary
conditions can be used for this purpose.
The final prefactor depends only on the ratio
$\chi(\tau_F)/\chi_0(\tau_F)$ and is therefore independent of the
specific choice of ${\tilde {\cal L}}_0$.
 
The function $\chi(\tau)$ satisfies Eq. (\ref{serve1}) while $\chi_0(\tau)$ 
solves the equation 
\beq
0 = {\tilde {\cal L}}_0 \, \chi_0(\tau) \; , 
\label{mistozero}
\eeq
where ${\tilde {\cal L}}_0$ is the reference operator. 
In general, as said before, the choice of this reference operator
is rather arbitrary,
        as any operator with a regular, known solution satisfying the
        same boundary conditions can be used; 
        the physical prefactor is determined by the ratio of the solutions
        at the final time, which is invariant under such a choice.
    
According to the Gel'fand-Yaglom theorem \cite{gy-method1,gy-method2},
the functions $\chi(\tau)$ and $\chi_0(\tau)$ must satisfy Dirichlet
boundary conditions at the lower endpoint: 
\beq
\chi(0)=\chi_0(0)=0 \; , \qquad \chi'(0)=\chi_0'(0) =1 \;  
\eeq
These conditions ensure the regularity and a unique normalization
of the determinant ratio.  
Physically, the condition $\chi(0)=0$ corresponds to vanishing fluctuations 
$\delta a(0)=0$ at the starting point of the instanton trajectory, while 
$\chi'(0)=1$ fixes the overall scale of the mode function.  
The reference operator ${\tilde {\cal L}}_0$ of Eq. (\ref{mistozero}) 
is chosen as 
\beq 
    {\tilde {\cal L}}_0 = -{d^2\over d\tau^2} + {1\over 4} \; ,
    \label{refoper}
\eeq
and its normalized solution, such that $\chi_0(0)=0$ and $\chi_0'(0)=1$,
is given by 
\beq
\chi_0(\tau)=2\sinh\left({\tau\over 2}\right) \; . 
\label{mino2}
\eeq
Notice that the reference operator (\ref{refoper}) is
    chosen for its constant coefficients and for the fact that its solution
    with Dirichlet boundary conditions, $\chi_0(\tau)$, is simple and
    regular. 
As previously stressed, the Gel'fand-Yaglom theorem \cite{gy-method1,gy-method2}
states that the determinant ratio, and hence the prefactor,
    is completely determined by
    $\chi(\tau_F)/\chi_0(\tau_F)$, independent of the absolute normalization
    of either $\chi(\tau)$ or $\chi_0(\tau)$. In this way, the functional
    determinant is correctly normalized and provides a physically meaningful
    estimate of the tunneling prefactor, while keeping the calculation
    analytically tractable.

The determination of $\chi(\tau)$ and $\chi(\tau_F)$ is much more complicated. 
It is useful to introduce the variable
\beq 
y=\cos\left({\tau\over \sqrt{3}}\right)
\eeq
and the function 
\beq
\psi(y) = \chi(\tau(y)) 
\eeq
such that
\beq
\chi(\tau_F) = \psi(y_F)
\eeq
with $\tau_F=\pi\sqrt{3}/2$ and $y_F=0$. 
In terms of $\psi(y)$, Eq.~(\ref{serve1}) becomes
\beq
0 = \left[ (1-y^2) {d^2\over dy^2} - y {d\over dy}
  + \frac{9}{4} - \frac{1}{4(1-y^2)}\right]\psi(y)  \; 
\label{eqpsi}
\eeq
where trigonometric functions do not appear explicitly.
The problem becomes a bit simpler setting 
\beq
\psi(y) = (1 - y^2)^{1/2} \, u(y) \; . 
\eeq
In this way, from Eq. (\ref{eqpsi}) we find 
\beq
0 = \left[(1-y^2) {d^2\over dy^2} - 2 y {d\over dy} + 2 \right] u(y) \; . 
\eeq
This is a Legendre equation, whose general solution reads \cite{gradshteyn}
\beq
u(y) = C_1 \, P_1^{1/2}(y) + C_2 \, Q_1^{1/2}(y) \; 
\eeq
with $P_{\nu}^{\mu}(y)$ the Legendre polynomials of the first kind
and $Q_{\nu}^{\mu}(y)$ the Legendre polynomials of the second kind.
Here $\nu =1$ and $\mu=1/2$.

Regularity at $\tau = 0$ (i.e. $y=1$) requires the solution to remain finite,
which eliminates the divergent term proportional to $P_1^{1/2}(y)$,
thus setting $C_1=0$.
The normalization $\chi'(0)=1$ then fixes $C_2$ uniquely. These
conditions imply that $u(y)$ is regular and properly normalized
at $y=1$ with 
\beq 
C_1=0 \quad\quad \mbox{and} \quad\quad 
C_2 = {8\sinh({\tau_F\over 2})
  \over \sqrt{\pi}}{\Gamma(3/4)\over \Gamma(5/4)} \; . 
\eeq
Thus, the compact final formula of $\chi(\tau)$ is 
\beq
\chi(\tau) = {8 \sinh({\pi \sqrt{3}\over 4})\over \sqrt{\pi}}
    {\Gamma(3/4)\over \Gamma(5/4)} \sin\left({\tau\over\sqrt{3}}\right) \,
    Q_1^{1/2}\left(\cos\left({\tau\over \sqrt{3}}\right)\right) \; . 
\eeq
The value of $Q_1^{1/2}(0)$ is known explicitly:
\beq
Q_1^{1/2}(0) = \frac{\sqrt{\pi}}{2} \frac{\Gamma(3/4)}{\Gamma(5/4)}.
\eeq
Then it follows that 
\beq
\chi(\tau_F) = \psi(y_F) = C_2 \, Q_1^{1/2}(0) =
4\sinh\left({\tau_F\over 2}\right) {\Gamma(3/4)^2\over \Gamma(5/4)^2} \; .
\label{mino3}
\eeq
In conclusion, from Eqs. (\ref{mino1}), (\ref{mino2}), and (\ref{mino3}) 
with $\tau_F=\pi\sqrt{3}/2$ (i.e., $y_F=0$),
we obtain the prefactor $F_E$ of Eq. (\ref{bastacosi}). 

\end{document}